  \documentclass[twocolumn,prl,aps]{revtex4}

\usepackage[dvips]{graphicx}
\usepackage[dvips]{graphics}

\begin{document}

 \title{Magnetic order in the pseudogap phase of high-$T_C$ superconductors}

 \author{B. Fauqu\'e$^1$, Y.~Sidis$^1$, V.~Hinkov$^2$, S.~Pailh\`{e}s$^{1,3}$, C.T. Lin$^2$, 
X. Chaud$^4$ and P.~Bourges$^{1,*}$}

\affiliation{
$^1$ Laboratoire L\'{e}on Brillouin, CEA-CNRS, CEA-Saclay, 91191 Gif sur Yvette, France\\
$^2$ MPI f\"ur Festk\"orperforschung, Heisenbergstr. 1, 70569 Stuttgart, Germany\\
$^3$ LNS, ETH Zurich and Paul Scherrer Institute, CH--5232 Villigen PSI, Switzerland\\
$^4$ CRETA / CNRS, 25 Avenue des Martyrs, BP 166 38042 Grenoble cedex 9, France.
}

\begin{abstract}
One of the leading issues in high-$T_C$ superconductors is the origin of the pseudogap 
phase in underdoped cuprates. Using polarized elastic neutron diffraction, we identify 
a novel magnetic order in the YBa$_2$Cu$_3$O$_{6+x}$ system. 
The observed magnetic order preserves translational symmetry
as proposed for orbital moments in the circulating current theory of the pseudogap state.
To date, it is the first direct evidence 
of an hidden order parameter characterizing the pseudogap phase in high-$T_C$ 
cuprates. 
\end{abstract}

\maketitle

In optimally and underdoped regimes, high-$T_C$ copper oxides superconductors exhibit 
a pseudogap state\cite{workshop,revue,timusk,tallon} with 
anomalous  magnetic\cite{alloul}, transport\cite{ito}, thermodynamic\cite{loram} 
and optical\cite{timusk} properties below a temperature, T$^*$, large compared to the
superconducting transition temperature, T$_c$.  The origin of the pseudogap is 
a challenging issue as it might eventually lead to identify the superconducting 
mechanism\cite{workshop}. Two major classes of theoretical models attempt to describe
the pseudogap state: in a first case, it represents a precursor of the superconducting 
$d$-wave gap\cite{rvb,preformedpairs} with preformed pairs below T$^*$
which would acquire phase coherence below T$_c$\cite{preformedpairs,orenstein}. 
In a second approach, the pseudogap is associated either with an ordered 
\cite{cmv-prb,simon,ddw,cdw,sdw,poilblanc} or a disordered phase 
\cite{workshop,stripes,fop} 
competing with the SC one.  The order parameter, associated with these competing phases
may involve charge and spin density waves\cite{cdw,sdw,poilblanc} or 
charge currents flowing around the CuO$_2$ square lattice, such as D-charge
density wave (DDW) \cite{ddw} or orbital circulating currents (CC) \cite{cmv-prb,simon}.

Most of these phases break the translation symmetry of the
lattice (TSL). Therefore, they may induce charge, nuclear or magnetic superstructures
that can be probed by neutron or X-ray diffraction techniques.
In contrast, CC phases\cite{cmv-prb,simon} preserve the TSL as they
correspond to 4 or 2 current loops per unit cell 
 (referred as $\Theta_{I}$ and $\Theta_{II}$ phases, respectively).
These charge currents could be identified by virtue of 
the pattern of ordered orbital magnetic moments pointing perpendicularly to the CuO$_2$ 
planes. These orbital magnetic moments should be detectable by neutron diffraction. 
Although the TSL is preserved, the magnetic signature of the 
CC phase does not reduce to ferromagnetism: the loops are 
staggered within each unit cell corresponding to a zero magnetic propagation wavevector, 
{\bf Q}=0, but with no net magnetization. In neutron diffraction, the magnetic intensity 
superimposes on the nuclear Bragg peak, meaning that these experiments are very delicate 
as the magnetic intensity $\propto M^2$ (M is the magnetic moment) is expected
to be very small as compared to the nuclear Bragg peaks. In order to detect this 
hidden magnetic response, polarized neutron experiments are then required. 

As proposed by C.M. Varma\cite{cmv-prb,simon}, there are two possibles CC phases 
preserving TSL. The first (the phase $\Theta_{I}$) has not been detected by polarized 
elastic neutron scattering experiments \cite{dhlee,bourges-lpr}. Although it is
controversial, a recent ARPES mesurement observed a dichroic signal in the 
$\rm Bi_2 Sr_2 Ca Cu_2 O_{8+\delta}$ system consistent with 
the phase  $\Theta_{II}$\cite{kaminski}. Here, we have performed polarized elastic 
neutron scattering 
experiments  to test the magnetic moments of this second CC state, phase $\Theta_{II}$, 
which actually had never been attempted before. We successfully report the first 
signature of a novel magnetic order in the pseudogap state of YBa$_2$Cu$_3$O$_{6+x}$ 
(YBCO). The pattern of the observed magnetic scattering corresponds to the one
expected in the circulating current theory of the pseudogap state with two current 
loops per CuO$_2$ unit-cell, phase $\Theta_{II}$ \cite{cmv-prb,simon}.
Alternatively, a decoration of the unit cell with staggered moments on the oxygen 
sites could also account for the measurements. 

All the polarized neutron diffraction measurements were collected on the 4F1 triple-axis 
 spectrometer at the Laboratoire L\'eon Brillouin (LLB), Saclay (France). 
Our polarized neutron diffraction setup is similar to that originally described 
in \cite{moon} with a polarized incident neutron with at $E_i=14.7$ meV obtained 
with a polarizing supermirror (bender) and with an Heusler analyzer 
(see also ref. \cite{yvan,dhlee} in the context of high-T$_C$ cuprates). 
The direction of the neutron spin polarization, {\bf P}, at the sample 
 position is selected by a small guide field {\bf H} of the order of 
10 G\cite{magnetic_scattering}.
Using that configuration, we monitor for each measured point 
the neutron scattering intensity in the spin-flip (SF) channel, where the magnetic 
intensity $\propto M^2$ is expected, and in the 
non-spin-flip (NSF) channel which measures the nuclear scattering. 
To have similar counting statistics on both SF and NSF, we count the SF channel 
systematically 20 times longer than the NSF. We define the normalized spin-flip intensity as 
$I_{norm} = I_{SF}/I_{NSF}$ (inverse of the flipping ratio (FR)). With that setup, 
a typical flipping ratio, ranging between 40 and 60, is obtained. However, even 
with that high FR, the SF intensity is massively coming from the 
NSF nuclear Bragg peak through unavoidable polarization leakage (corresponding 
to about $\sim$ 90-95\% of the SF intensity). As a very stable and homogeneous 
neutron polarization is essential through 
the data acquisition, all the data have been obtained in a continuous run versus 
temperature. We prove 
that method to be efficient enough to see weak magnetic moments ($\sim 0.05 \mu_B$) 
on top of nuclear Bragg peaks, see e.g. the first determination of the A-type 
antiferromagnetism in Na cobaltate systems\cite{sibel}. 

We quote the scattering wave vector as $\bf{Q}$=(H,K,L) in units of the reciprocal lattice 
vectors, $a^* \sim b^*$ = 1.63 \AA$^{-1}$ and $c^*$ = 0.53 \AA$^{-1}$.
Most of the data have been obtained in a scattering plane where all Bragg peaks like 
$\bf{Q}$=(0,K,L) were accessible (in twinned samples, this is indistinguishable 
from Bragg peaks with $\bf{Q}$=(H,0,L)). In order to evidence small magnetic moments, 
measurements have been performed on the weakest nuclear Bragg peaks having the proper
symmetry for the CC phase\cite{structurefactor} (the Bragg peak $\bf{Q}$=(0,1,1) offers the best compromise).
 
\begin{table}[t]
\begin{center}
\begin{tabular}{|c|c|c|c|c|}
\hline
label & x & T$_{c,onset}$ (K) & T$_{mag}$(K) & References\\
\hline
A & ${\rm O_{6.5} (t)}$ & ud 54 & 300 $\pm$ 10 & \cite{yvan}\\
\hline
B & ${\rm O_{6.6} (t)}$ & ud 61 & 250 $\pm$ 20 &  \cite{lothar}\\
\hline
C & ${\rm O_{6.6} (d)}$ & ud 64 & 220 $\pm$ 20 & \cite{hinkov}\\
\hline
D & ${\rm O_{6.75} (t)}$ & ud 78 & 170 $\pm$ 30 & -\\
\hline
E & ${\rm Ca(15\%)-O_{7-\delta} (t)}$ &  od 75 & $\simeq$0 & -\\
\hline
 \end{tabular}
\caption{\label{table1}List of samples utilized in the polarized elastic neutron experiments. 
The experiments were performed in the (Y,Ca)Ba$_2$Cu$_3$O$_{6+x}$ family for 5 samples
from the underdoped (ud) to overdoped (od) part of the cuprates phase diagram. 
(t) and (d) stands for twinned and detwinned samples, respectively. References
are given where the samples have been described in previous neutron scattering 
studies. In contrast to the other samples, an in-plane magnetic ordering 
occurred at ${\bf Q}$=(1/2,1/2) in sample A with  $M\sim 0.05 \mu_B$ at 60 K\cite{yvan}. }
\end{center}
\end{table}

\begin{figure}
 \includegraphics[height=10cm,width=8cm]{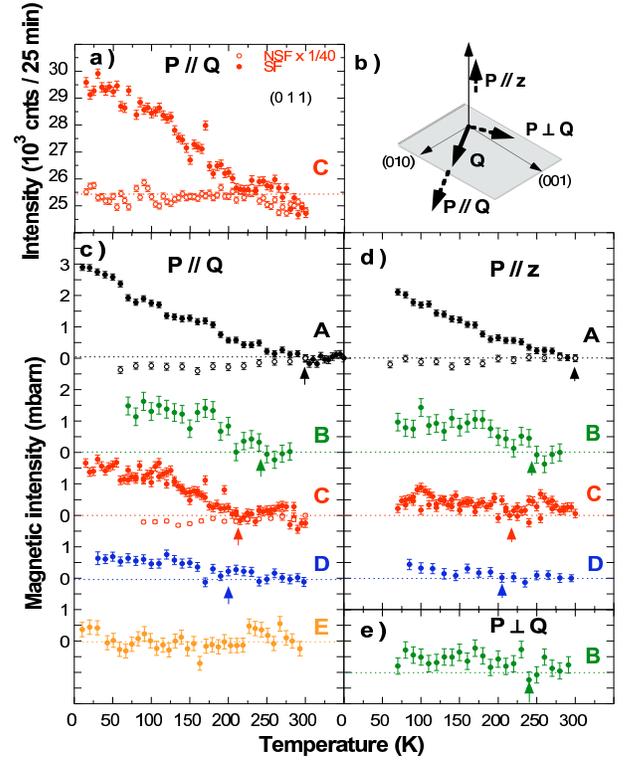}
\begin{center}
\caption{(color online) (a): Temperature dependencies of the raw SF and NSF neutron intensity measured 
at $\bf{Q}$=(0,1,1) in sample C.
(b) Sketch of the scattering plane showing the three polarization directions discussed 
here, ${\bf P}$//z corresponds to the direction perpendicular to the scattering plane
(here a$^*$). c) Temperature dependencies of the normalized magnetic intensity, $I_{mag}$, measured at 
$\bf{Q}$=(0,1,1) for $\bf{P}$//$\bf{Q}$ for the 4 underdoped 
samples (A,B,C,D) and the overdoped sample E (full points). $I_{mag}$ is defined as 
$I_{mag}(T)= \alpha I_{NSF}(300 K) \big[ \frac{I_{SF}}{I_{NSF}}(T)- \frac{I_{SF}}{I_{NSF}}(T \sim 300 K) \big ]$
where (i) $I_{mag}$ is arbitrarily set to zero in the high temperature range ($\sim$ room temperature), 
(ii) $\alpha=7/I^{meas}_{004}(300 K)$ calibrates the magnetic cross-sections in 
mbarns using the nuclear Bragg cross-section at ${\bf Q}=(0,0,4)$, $I^{calc}_{004}$= 7 barns. 
The normalized magnetic intensity for the Bragg peak, ${\bf Q}=(0,0,2)$, is also 
shown for samples A and C (open points). d)  Temperature dependencies of the 
normalized magnetic intensity measured at $\bf{Q}$=(0,1,1) 
(full points) (as well as $\bf{Q}$=(0,0,2), open points) for $\bf{P}$//$\bf{z}$.  
e) Temperature dependencies of the 
normalized magnetic intensity, $I_{mag}$, measured at $\bf{Q}$=(0,1,1) 
for sample B for $\bf{P}\perp\bf{Q}$.   } 
\label{fig1}
\end{center}
\end{figure}

We have studied 5 different samples (see Table \ref{table1}): 4 samples in 
the underdoped regime and one in the overdoped regime. In Fig. \ref{fig1}.a, 
we report the raw neutron intensity measured at $\bf{Q}$=(0,1,1) for the spin 
flip (SF) channel  and  for the non-spin-flip (NSF) channel 
for an underdoped sample YBa$_2$Cu$_3$O$_{6.6}$(d) (sample C). 
The measurement has been done with a neutron polarization  $\bf{P}$//$\bf{Q}$ (see Fig. \ref{fig1}.b)
where the magnetic scattering is entirely spin-flip\cite{moon,dhlee,yvan,magnetic_scattering}. 
Between room temperature 
and a temperature T$_{mag}$$\simeq$220K, the NSF and SF intensities display the 
same evolution within error bars. Then, for T$<$T$_{mag}$, the NSF is essentially flat  whereas the SF 
intensity increases noticeably at low temperature. This behaviour signals the 
presence of a spontaneous magnetic order below T$_{mag}$ on top of the nuclear 
Bragg peaks. In Fig. \ref{fig1}.c, we show the normalized magnetic intensity 
as a function of the temperature for the 4 underdoped samples and the overdoped 
sample. For the 4 underdoped samples, the magnetic intensity increases at low 
temperature below a certain temperature T$_{mag}$ whereas no magnetic signal is 
observed in the Ca-YBCO overdoped sample (sample E). 

We  perform further measurements where the neutron polarization 
is along the complementary directions, as shown in Fig. \ref{fig1}.b, either the 
vertical direction ${\bf P}//{\bf{z}}$, or ${\bf P}\perp{\bf{Q}}$ but still within the 
horizontal scattering plane. The observance of the polarization selection
rule for a magnetic signal, $I_{{\bf P}//{\bf Q}}=I_{{\bf P}//{\bf z}}+I_{{\bf P}\perp{\bf Q}}$,
in the three polarizations, as shown in Fig. \ref{fig1}.c,\ref{fig1}.d 
and \ref{fig1}.e for sample B, unambiguously demonstrates the magnetic origin
of the low temperature signal. 
More precisely, in the ${\bf P}//{\bf{z}}$,  configuration, only 
magnetic moments within the horizontal scattering plane but still perpendicular 
to ${\bf Q}$ are observed in SF channel \cite{moon,dhlee,yvan,magnetic_scattering}. 
For ${\bf Q}=(0,1,1)$, this 
means that we mostly probe the magnetic moments parallel to the c$^*$ axis. In 
the 4 underdoped samples, we observe a similar onset of the magnetic order below 
$T_{mag}$ for $\bf{P}$//$\bf{z}$ (Fig. \ref{fig1}.d). This demonstrates that the 
deduced magnetic moment has a well-defined component perpendicular to the CuO$_2$ plane. 
However, a closer comparison with both polarizations reveals that their 
intensities do not simply match. This underlines that the magnetic moment also exhibits
an in-plane component (within the CuO$_2$ plane) as the cross-section in 
Fig. \ref{fig1}.c is larger than the one in Fig. \ref{fig1}.d. 
Combining all measured polarizations in the different samples, one can then  
estimate a mean angle between the direction of the moments with the c$^*$ axis to 
be $\phi=45^{\circ} \pm 20^{\circ}$ valid for all samples. 

\begin{figure}[t]
 \includegraphics[height=4.5cm,width=8cm]{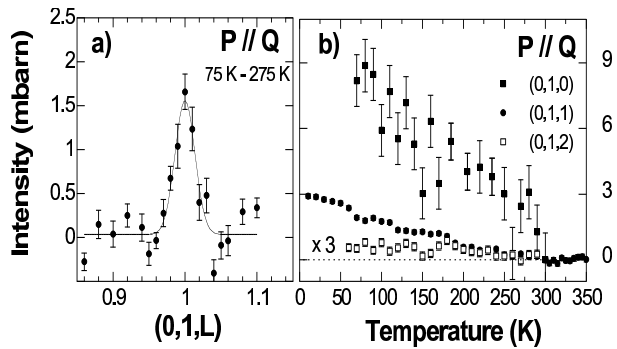}
\begin{center}
\caption{a) L-scan magnetic intensity across Q=(0,1,L) in sample A: it has been obtained 
using the following relation of measured quantities 
$\Large[ I(SF,75K)-I(SF,275K) \frac{I(NSF,75K)}{I(NSF,275K)} \Large]$ calibrated by
$\alpha$ (see caption Fig. \ref{fig1}).
b) Temperature dependencies of the magnetic intensity, $I_{mag}$, for various
Bragg peaks  L=0,1,2 in sample A.} \label{fig2}
\end{center}
\end{figure}

As shown in Fig. \ref{fig1}.c, the typical cross-section of the 
magnetic order is $\sim$ 1--2 mbarns, {\it i.e.} $\sim$ 10$^{-4}$ of the strongest 
Bragg peaks. This explains why such a magnetic order was not reported before 
with unpolarized neutron diffraction. Due to these experimental limitations,
we do not perform a detailed and quantitative determination of magnetic structure 
for which further work is needed. However, some qualitative aspects can be briefly 
discussed. First, we perform a scan along the L-direction in the SF channel across 
the Bragg peak (Fig. \ref{fig2}.a) where the difference in temperature between T=75 K 
and 275 K has been taken to remove the effect of the polarization leakage.
The observed magnetic peak is 
resolution limited, showing that the magnetic  order is characterized by long range 3D  
correlations at T=75 K. Second,  by looking at other 
Bragg peaks along c$^*$ (Fig. \ref{fig2}.b), we found that the magnetic intensity 
is not uniformly distributed versus $L$, meaning that i) the magnetic intensity does 
not arise from the Cu-O chains, and ii) the moments arrangement within a bilayer 
appears to be mainly parallel. This directly arises from the hierachy of the observed 
magnetic intensities (intensity at $L$=0 is larger than at $L$=2, Fig. \ref{fig2}.b).
Finally, using the observed magnetic cross-section (Fig. \ref{fig1}.c) and a weakly 
momentum dependent form factor, one can deduce a typical magnitude of ordered magnetic
 moment of $M \simeq 0.05$ to $0.1 \mu_B$ with the moment decreasing with increasing 
doping in the 4 samples. 

\begin{figure}[t]
 \includegraphics[width=8cm]{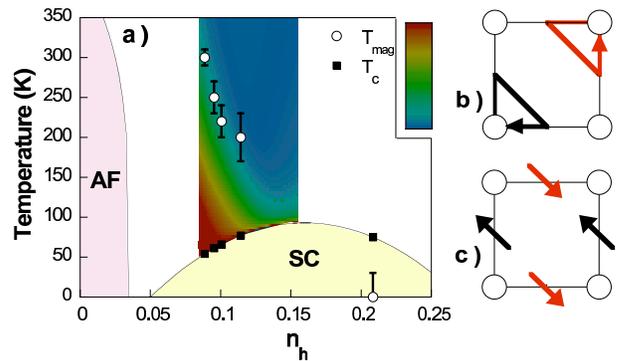}
\begin{center}
\caption{(color online) a) Cuprate superconductors phase diagram as a function of hole doping, $n_h$,
deduced from the SC temperature using the empirical relation 
$T_C/T_C^{max}=1-82.6(n_h-0.16)^2$\cite{tcnh}. The white points show $T_{mag}$ 
(see table \ref{table1}).  
The color map shows the quantity $\delta R(T)=1-[\rho_{ab}(T)-\rho_{ab}(0)]/(\alpha T)$ 
deduced from the resistivity measurements in YBCO\cite{ito}: 
the change of colors indicates the departure from the T-linear behaviour, $\delta R(T)= 0$  represented 
in blue, at high temperature. $\delta R(T) \ne 0$ defines 
the pseudogap state. b) Circulating current phase, $\Theta_{II}$, in the CuO$_2$ plane 
proposed to explain the pseudogap phase in high-$T_C$ superconducting cuprates\cite{cmv-prb,simon}.
c) A spin model preserving TSL. } \label{fig3}
\end{center}
\end{figure}

Therefore, we observe an unusual magnetic order in a temperature and doping 
range that cover the range where the pseudogap state is observed in YBCO. 
Our data do not contradict previous unsuccessfull polarized neutron reports\cite{dhlee}
as the Bragg spots where the effect is observed are along a direction at 45$^{\circ}$
from the one previously studied\cite{structurefactor}. The deduced $T_{mag}$, defined as the change of slope in the normalized 
intensity  $I_{mag}$, decreases with increasing doping (see table \ref{table1}). 
It  matches the pseudogap temperature, T$^{*}$, of the resistivity data 
in YBCO\cite{ito} as shown on Fig. \ref{fig3}. 
The occurrence of a magnetic order in this temperature and doping ranges points
towards a magnetic signature of an hidden order parameter associated with the pseudogap state.  As all anomalous physical properties evidencing the pseudogap, the temperature dependence 
of the magnetic order does not exhibit a marked change at $T_{mag}$. Being on top of nuclear Bragg peaks, that magnetic order does not break TSL, 
indicating a zero magnetic propagation wavevector, ${\bf Q}=0$. 
As shown in Fig. \ref{fig1}.c, no magnetic intensity occurs below $T_{mag}$ 
at the Bragg peak ${\bf Q}=(0,0,2)$, ruling out a 
ferromagnetic order. The absence of breaking of TSL points towards a 
magnetic pattern of antiparallel magnetic moments within each unit cell.
Among the proposed order parameters, only one gives 
magnetic scattering at  ${\bf Q}=(10L) [\equiv (01L)]$: it is the orbital moments 
arising from the circulating current phase with 2 current loops per 
CuO$_2$ unit cell, $\Theta_{II}$ (Fig. \ref{fig3}.b) \cite{cmv-prb,simon}.
Another possibility could be a model with colinear spin moments located at the oxygen site
ferromagnetic along both directions of the square lattice but antiparallel each other 
as sketched on Fig. \ref{fig3}.c. Any other model characterized by a decoration of the 
CuO$_2$ plaquette would also give rise to a magnetic contribution at 
the proper Bragg spots. 

From our present measurements, one cannot distinguish between these two models.
Only a detailed study of magnetic form factors would allow to 
differentiate the scattering from spin and orbital moments. However, some 
arguments can be given from the observed moments direction, which is
not within the  CuO$_2$ plane neither perpendicular to it.
Clearly, an in-plane magnetic component is not expected within the orbital moments 
picture of currents flowing within perfectly flat CuO$_2$ planes. 
However, due to the dimpling of CuO$_2$ planes in YBCO, the moments 
can be tilted by about 11$^{\circ}$ from the c axis as the effective 
moments at the centers of 
the O-Cu-O plaquettes are perpendicular to these plaquettes. Within an orbital moment 
picture, spin degree of freedom might also play a role in 
producing in-plane magnetic moments, for instance, by spin-orbit scattering\cite{so} or in
relation to chiral spin states associated with flux phases \cite{chiral}. 
Alternatively, if considering spin models, one would rather expect moments lying within 
the CuO$_2$ plaquette as it is the case for copper spins in undoped cuprates. A reason should 
be found to explain why the moments exhibit an out-of-plane component. 
Whatever the origin of the observed order, its pattern challenges the 
single band Hubbard 
picture commonly used to describe high-$T_C$ cuprates. At very least, 
oxygen orbitals need to be included to determine the minimal effective 
Hamiltonian for the cuprates. 

In conclusion, we report a first signature of an unusual magnetic order in several YBa$_2$Cu$_3$O$_{6+x}$ 
samples matching the pseudogap behaviour in underdoped cuprates (Fig. \ref{fig3}). 
Such an observation points towards the existence of an 
hidden order parameter for the pseudogap phase in high-$T_C$
superconductors. Importantly, our experiment reveals that a 3D long range order
does not break the translational symmetry of the lattice and implies a
decoration of the unit cell with staggered spin or orbital moments.
The symmetry of the observed order 
corresponds to the one expected in orbital moments emanating from a circulating 
current state \cite{cmv-prb,simon}. 

We are very grateful to C.M. Varma for invaluable encouragement, critics and ideas on 
these experiments. We also thank B. Keimer, J.-M. Mignot, P. Monceau, 
L. Pintschovius, and L.-P. Regnault for their support.

$^*$To whom correspondence should be addressed;
 E-mail: bourges@llb.saclay.cea.fr

 \clearpage 


\begin{thebibliography}{99}

\bibitem{workshop}  M.R. Norman, D.P. Pines, \& C. Kallin, preprint
cond-mat/0507031.

\bibitem{revue} M.R. Norman \& C. P\'epin, {\it Rep. Prog. Phys.} \textbf{66}, 1547 (2003). 

\bibitem{timusk} T. Timusk,  \& B. Statt, 
{\it Rep. Prog. Phys.} \textbf{62}, 61 (1999).

\bibitem{tallon} J.L. Tallon \& J.W. Loram,  {\it Physica C} {\bf 349}, 53 (2001).

\bibitem{alloul} H. Alloul {\it et al.} 
{\it Phys. Rev. Lett.} \textbf{63}, 1700 (1989).
  
\bibitem{ito} T. Ito {\it et al.} 
{\it Phys. Rev. Lett.} \textbf{70}, 3995 (1993).

\bibitem{loram} J.W. Loram  {\it et al.} 
{\it Physica C},  {\bf 235-240} 134 (1994).

\bibitem{rvb} P.A. Lee, {\it Physica C} {\bf 317--318}, 194--204, (1999).

\bibitem{preformedpairs} V.J. Emery \& S.A. Kivelson, {\it Nature} {\bf 374}, 
434 (1995).

\bibitem{orenstein} J. Orenstein, \& A.J. Millis, 
{\it Science} \textbf{288}, 468 (2000).

\bibitem{cmv-prb} C.M. Varma, {\it Phys. Rev. B}, \textbf{55}, 14554 (1997);
 {\it Phys. Rev. Lett.} \textbf{83}, 3538 (1999); preprint, cond-mat/0507214.

\bibitem{simon} M.E. Simon \& C.M. Varma, {\it Phys. Rev. Lett.} \textbf{89}, 247003 (2002).

\bibitem{ddw} S. Chakravarty  {\it et al.} 
{\it Phys. Rev. B} \textbf{63}, 094503   (2001). 

\bibitem{cdw} C. Castellani  {\it et al.} 
  {\it Phys. Rev. Lett.} \textbf{75}, 4650 (1995).

\bibitem{sdw} H.C. Chen,  {\it et al.} 
{\it Phys. Rev. Lett.} {\bf 93}, 187002 (2004).

\bibitem{poilblanc} D. Poilblanc,  cond-mat/0503249.
 

\bibitem{stripes} J. Zaanen,  {\it et al.} 
{\it Phil. Mag. B}, {\bf 81}, 1485 (2001).

\bibitem{fop}  F. Onufrieva, P. Pfeuty,
{\it Phys. Rev. Lett.}, {\bf 82}, 3136 (1999).

\bibitem{dhlee} S.H. Lee {\it et al.} 
{\it Phys. Rev. B} \textbf{60}, 10405 (1999). 

\bibitem{bourges-lpr} Ph. Bourges, L.P. Regnault, J.Y. Henry,  \& C. Marin, 
Unpublished data (1998). 

\bibitem{kaminski} A. Kaminski,  \emph{et al.}, 
{\it Nature} \textbf{416}, 610 {(2002);}
 {S. Borisenko,}  \emph{et al.}, {\it Nature} \textbf{431}, {(2 September 2004);}
 A. Kaminski,  \emph{et al.}, {\it ibid}.

\bibitem{moon}  R.M. Moon  {\it et al.} 
{\it Phys. Rev.} 181, 920 (1969).

 \bibitem{yvan} Y. Sidis,  {\it et al.} {\it Phys. Rev. Lett.} {\bf 86}, 4100 (2001).

\bibitem{magnetic_scattering} Magnetic neutron diffraction always measures 
magnetic components perpendicular to the scattering wavevector. In a polarized 
experiment, only the magnetic components perpendicular to the neutron polarization 
direction contributes to the spin-flip channel\cite{moon}.

\bibitem{sibel}  S.P. Bayrakci {\it et al.} {\it Phys. Rev. Lett.} \textbf{94}, 157205 (2005).

\bibitem{structurefactor} Bragg magnetic peaks characteristic of the two CC states 
proposed\cite{simon} differ by 45$^{\circ}$: main Bragg peaks like ${\bf Q}=(11L)$ 
are expected for the state $\Theta_{I}$ and like ${\bf Q}=(10L)$ [$\equiv (01L)$] 
for the state $\Theta_{II}$. In both cases, no magnetic contribution 
occurs on Bragg peaks like ${\bf Q}=(00L)$.

\bibitem{so} C. Wu, Y. Zaanen \& S.C. Zhang, cond-mat/0505544.

\bibitem{chiral} X.G. Wen  {\it et al.} 
 {\it Phys. Rev. B}, {\bf 39}, 11413 (1989).

\bibitem{hinkov} V. Hinkov {\it et al.}, {\it Nature} {\bf 430}, 650 (2004).

\bibitem{lothar} L. Pintschovius {\it et al.},
{\it Phys. Rev. Lett.} \textbf{89}, 037001 (2002).

\bibitem{tcnh} J.L Tallon {\it et al.}, 
{\it  Phys. Rev B}, {\bf 51}, R12911 (1995).

 \end{thebibliography}
\end{document}